\documentclass{iucrjournals}
\usepackage{amsmath}
\usepackage{float}
\usepackage{rotating}
\usepackage{subcaption}

\title{Data Driven Drift Correction For Complex Optical Systems}

\author{Aashwin Mishra\IUCrCemaillink{aashwin@slac.stanford.edu}}
\author{Matt Seaberg\IUCrCemaillink{seaberg@slac.stanford.edu}}
\author{Ryan Roussel\IUCrCemaillink{rroussel@slac.stanford.edu}}
\author{Sanghoon Song\IUCrCemaillink{sanghoon@slac.stanford.edu}}
\author{Auralee Edelen\IUCrCemaillink{edelen@slac.stanford.edu}}
\author{Daniel Ratner\IUCrCemaillink{dratner@slac.stanford.edu}}
\author{Apurva Mehta\IUCrCemaillink{mehta@slac.stanford.edu}}

\affil{SLAC National Laboratory, 2575 Sand Hill Rd, Menlo Park, CA 94025, USA}

\begin{document} 
\maketitle 


\begin{abstract}
To exploit the thousand-fold increase in spectral brightness of modern light sources, increasingly intricate experiments are being conducted that demand extremely precise beam trajectory. Maintaining the optimal 
trajectory over several hours of an experiment with the needed precision necessitates active drift control. Here, we outline Time-Varying Bayesian Optimization (TVBO) as a data driven approach for robust drift correction, and illustrate its application for a split and delay optical system composed of six crystals and twelve input dimensions. Using numerical simulations, we exhibit the application of TVBO for linear drift, non-smooth temporal drift as well as constrained TVBO for multi-objective control settings, representing real-life operating conditions. This approach can be easily adapted to other X-ray beam conditioning and guidance systems, including multi-crystal monochromators and grazing-incidence mirrors, to maintain sub-micron and nanoradian beam stability over the course of an experiment spanning several hours.
\end{abstract}


\section{Introduction}
The increasing brightness of light sources, such as the diffraction-limited enhancement of the Advanced Photon Source (APS) and the high-repetition-rate enhancement of the Linac Coherent Light Source (LCLS), paves the way for a deeper understanding of fundamental processes at the heart of chemistry, biology, and materials sciences. However, these insights necessitate increasingly intricate experiments that demand extreme precision of beam alignment and stability over extended periods \cite{Schoenlein2016, Margraf2023}. For instance, experiments conducted at LCLS-II-HE will require the X-ray beam to maintain a diameter of just a fraction of a micron, with a pointing stability of a few nanoradians at the conclusion of a kilometer-long electron accelerator, a hundred-meter-long undulator section, and greater than one hundred meters of mirror and crystal-based X-ray transport. This configuration needs to be maintained for the entire experiment duration of the order of many hours.



Temporal drifts of beam trajectory can occur in X-ray source points and optical systems due to many factors, such as thermal variations, mechanical vibrations, and environmental changes. Such drift can affect the quality of the data generated. For illustration, due to their short wavelength, X-rays are often used to probe matter at the nanometer scale and as a result x-ray beams must often be focused to sub-micron size. Many experiments at the ultra bright X-ray facilities, including picosecond X-ray Photon Correlation Spectroscopy (XPCS) and Transient grating Spectroscopy, both of which depend on multiple beams maintaining a high degree of overlap on a sample, demand a high degree of beam position and pointing stability. As such, the beams must have nanoradian level stability on the timescale of the measurement in order to prevent beams losing overlap or a shift to a different region of a heterogeneously evolving sample. This level of stability is typically difficult to meet for a variety of reasons, including thermal/environmental stability and opto-mechanical imperfections. When the required stability can't be met and when it is not possible to correct for drift parasitically, measurements must be interrupted frequently to correct for errors that compromise data quality, resulting in significant data-collection dead-time and poorer quality measurements. 


There are many examples of drift correction techniques and algorithms developed to compensate for time-dependent trajectory alterations, both for lasers and for x-ray sources. These typically rely on a traditional feedback system, using either PID loops or in some cases neural networks \cite{Breitling2001, Genoud2011}. A critical aspect of such feedback systems is that they must have sufficient diagnostics such that the system can be diagonalized. This can be relatively straightforward for typical laser systems and even for synchrotron x-ray sources \cite{Muller2012,Martin2022}. In some cases, especially for high-powered lasers or x-ray sources, low-power optical guide beams can be used as a surrogate for the beam of interest \cite{Burkhart2011,Koehlenbeck2025}. However, in situations where guide beams are not available, and in which the system has more degrees of freedom than the number of independent diagnostic measurements such that traditional feedback can not be used, alternative approaches must be considered.

In this investigation, we outline the use of Time Varying Bayesian Optimization (TVBO) as an approach for drift correction in complex optical systems \cite{Kuklev2022,Kuklev2023,Xu2023}. We utilize TVBO with a fixed forgetting window approach and apply this for drift correction in the  Hard X-Ray Split and Delay system (HXRSND) at LCLS \cite{Zhu2017}. With the introduction of LCLS-II-HE at SLAC, the HXRSND  will play a pivotal role in investigations of complex materials using tools for ultrafast XPCS and transient grating spectroscopy measurements \cite{Gutt2009}. Consequently, it is imperative that we prevent operational inefficiencies from affecting scientific throughput. We report successful application of TVBO for different cases including constant linear drift, non-smooth discontinuous drift and constrained TVBO for multi-objective control settings.

\section{Methods \& Application}

\subsection{Time Varying Bayesian Optimization}

Bayesian Optimization \cite{frazier2018bayesian, Roussel2024, edelen2019machine, mishra2025start} (BO) is a sequential sampling approach for finding global optima of black-box functions that are expensive to evaluate, noisy, or have uncertain dynamics, etc. It uses a surrogate probabilistic model (often using Gaussian Processes \cite{rasmussen2006gaussian}) that estimates the distribution of possible function values at points in the domain. An acquisition function is used to determine the next point to be sampled at, based on the predictions of the surrogate probabilistic model. The acquisition function balances exploration (i.e., a preference towards points where the surrogate probabilistic model's predictions have high variance) and exploitation (i.e., a preference towards points that have better mean predictions according to the model). Common acquisition functions include Expected Improvement (EI), Upper Confidence Bound (UCB), and Probability of Improvement (PI) \cite{garnett2023bayesian, snoek2012practical}. This sampling is carried out iteratively, where the 
point is evaluated, the surrogate probabilistic model is retrained with this augmented dataset, and the subsequent point is recommended based on the trained surrogate model's predictions and the appropriate acquisition strategy. The key advantage of Bayesian Optimization is that it allows optimization in minimal evaluations of the underlying process. This is useful both for time-consuming simulations as well as beamtime, where real-time feedback and decision latency are of critical importance. Secondly, it enables us to handle general black box functions without assuming any functional form, via the use of flexible Bayesian models like Gaussian Processes. Finally, owing to the use of probabilistic surrogate models, Bayesian Optimization is robust to noise in the function evaluations (e.g. the noise inherent to minimally-invasive diagnostics). In this vein, Bayesian Optimization has been used for many complex applications such as tuning of particle accelerators.

A limitation of traditional Bayesian Optimization is its assumption of the underlying objective function remaining static. Time Varying Bayesian Optimization (TVBO) \cite{brunzema2022controller} is an extension that addresses problems where the objective function changes over time. This can occur in many problems, for instance, dynamic environments, where the underlying system being optimized exhibits temporal variations due to factors like drift, seasonality, trends, or external influences. The utilization of TVBO for drift correction represents a step beyond classical use of feedback loops for drift correction (like PID controllers) by not just correcting a single variable to a fixed setpoint, but by actively re-optimizing the machine's performance in an evolving multi-dimensional parameter space as the operating conditions and the machine itself slowly change. Additionally, the Gaussian Process surrogate model in TVBO enables robustness to noisy measurements, which are ubiquitous in beamline optics. It learns the underlying function despite stochastic fluctuations.

The central challenge of TVBO is modifying the standard Bayesian Optimization framework, which assumes a static objective function, to handle a time varying function. In the recent past, multiple algorithms have been proposed to achieve this, that differ in the manner by which they forget old information, and how they model time dependency. This includes data-centric approaches \cite{pmlr-v51-bogunovic16, zhou2021no} like sliding window based TVBO, TVBO with weighing of samples; model-centric approaches \cite{liu2021gaussian} like utilizing a Gaussian Process model with a time-dependent kernel, utilizing dynamic Gaussian Process models; etc. In this investigation, we utilize Time Varying Bayesian Optimization with a sliding window approach \cite{zhou2021no} implemented in Xopt \cite{roussel2023xopt}. Here, the surrogate model is trained on the most recent $w$ samples. This enables the optimization procedure to focus on a recent subset of observations, which may be useful in dynamic environments where significantly older data may no longer be relevant and may even skew the optimization process. The fixed sized window of $w$ recent samples slides with every subsequent sample. The size of this window controls how quickly the surrogate model can adapt to changes in the objective function.




\subsection{Overview of the HXRSND}
\begin{figure}[h!]
\centering
\includegraphics[width=1.02\textwidth]{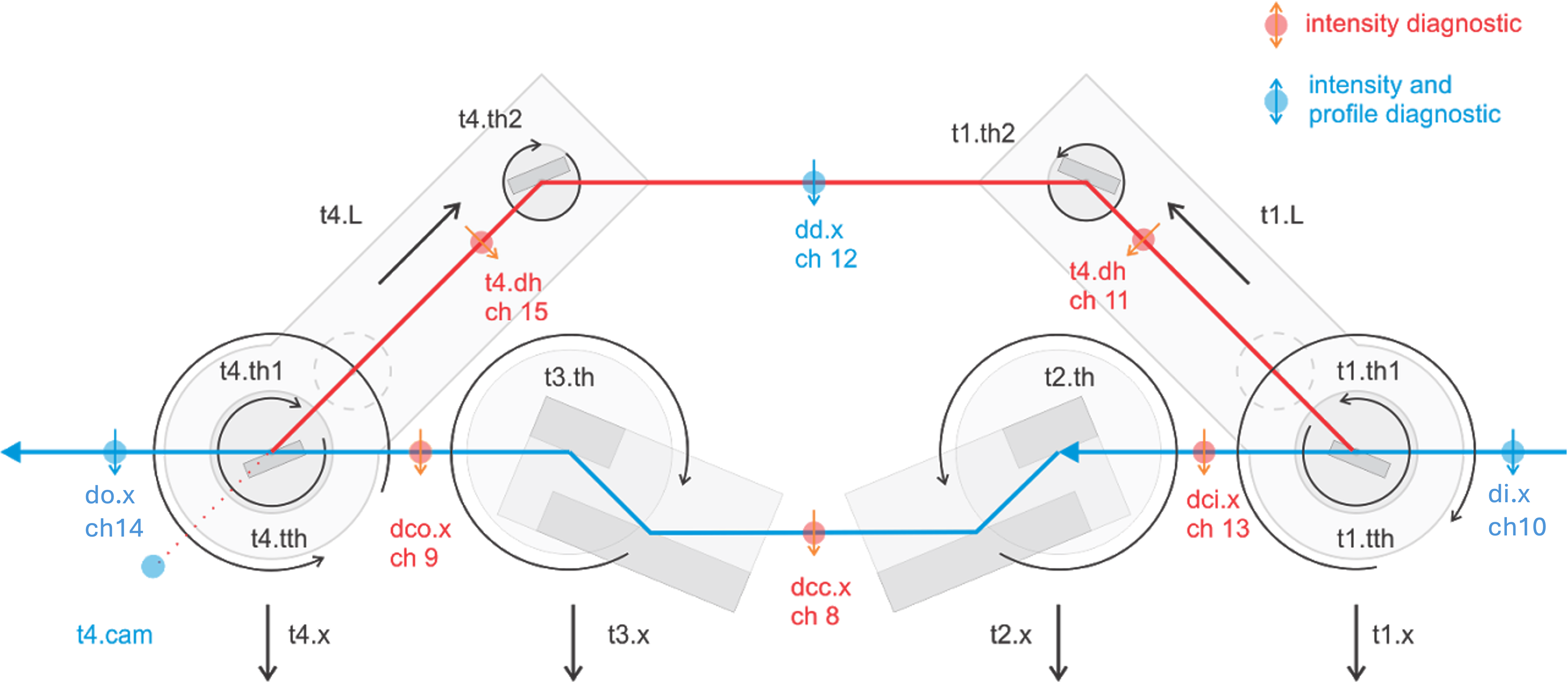}
\caption[]{Schematic of the HXRSND with the CC (Channel Cut) branch in blue and the delay branch in red. The x-ray beam propagates from right to left. The arrows indicate the motorized degrees of the system. In addition, each crystal along the delay branch has ``chi'' adjustment which corresponds to rotation about the tangential vector of the crystal surface (not shown). The red and blue dots correspond to locations of beam diagnostics, as noted in the legend on the top right of the figure. Figure re-used with permission from \cite{Zhu2017}.
}
\label{fig:fig1}
\end{figure}


In this investigation, we utilize the HXRSND as the system under study for TVBO. As depicted in Figure 1, this system comprises two branches: the minimally adjustable 'channel-cut' (CC) branch (represented in blue), and the 'delay' branch (illustrated in red), which possesses twelve degrees of freedom used to introduce a delay relative to the channel cut branch while maintaining a constant trajectory at the system output. The delay range spans from approximately -5 to 500 ps, corresponding to the path length difference between the branches ranging from -1.5 to 150 mm. The alignment of the HXRSND necessitates a spatial overlap between the two branches at the sample with exceptional precision, as well as optimized intensity at the output. For instance, for many experiments, both branches must be aligned to the same photon energy within approximately 0.1 eV with matching intensity, while simultaneously overlapping the resulting beams to a small fraction of the focused beam size.


For measurement techniques such as ultrafast XPCS, the two beams must maintain overlap at the $\sim 1$ micrometer level for many hours, while maintaining a constant path length difference of several millimeters, to maintain data quality \cite{Li2021}. Since the stability of the system is not sufficient for this, the overlap must be manually checked, and if necessary corrected, every $\sim 10$ minutes, forcing interruption of data collection frequently, requiring a constant operator presence and reducing data-collection efficiency. Typically this manual check involves the invasive insertion of a fluorescent YAG screen, but in principle potentially non-invasive measurements such as speckle from a static sample can provide the same information \cite{Li2021}. Furthermore, the system is relatively complex from an x-ray optics standpoint, such that when correction is performed manually the optical element used for correction may not correspond to the element that caused the drift. Over time, this approach leads to sub-optimal correction and can lead to a need for a more detailed re-alignment of the system.

\subsection{Details of the simulations}
\begin{figure}
        \centering
        \includegraphics[width=\textwidth]{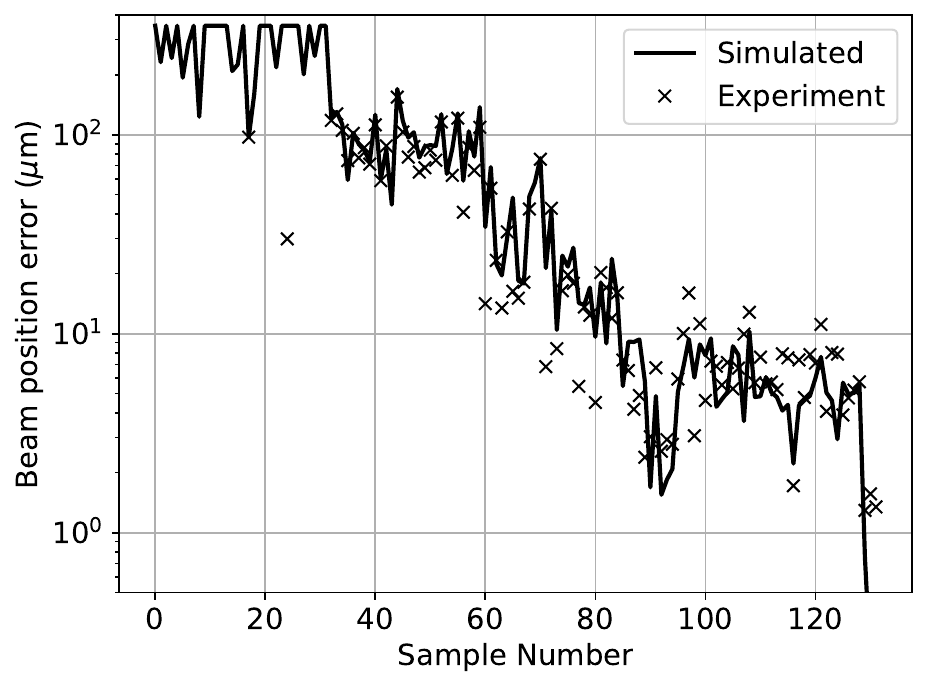}
        \caption{Comparison of the predictions of the Beam Position Error from the wave-optical simulations code used in this investigation against experimental results on the HXRSND (delay branch only). Between each sample various degrees of freedom were adjusted both in experiment and also as inputs to the simulator.}
        \label{fig:simulation_validation}
\end{figure}

For this study, we performed wave-optical simulations of the system in preparation of the limited availability of XFEL beamtime. The simulations are based on decoupled horizontal and vertical (2$\times$1D) wavefront propagation using in-house software\cite{lcls_beamline_toolbox}. These simulations model the input beam as fully spatially coherent (a reasonable assumption for XFEL beams) and monochromatic (taking credit for a monochromator upstream of the split and delay system). The various motion degrees of freedom shown in Figure~\ref{fig:fig1} are all reproduced in the simulation, and care was taken to ensure that the simulation sufficiently captures the dynamics of the operation of the actual HXRSND (see Figure~\ref{fig:simulation_validation}). Since the CC branch is intrinsically much more stable than the delay branch, we focus the simulation efforts on maintaining the stability of the delay branch. In this study, the system was configured for operation at 9.5 keV with zero relative delay between the branches. To judge spatial overlap, we simulate the position of the beam directly at the interaction point as if it were measured using YAG fluorescence. Even though this represents an invasive measurement, the results of the simulation study also apply to non-invasive measurements that are under development, assuming they provide equivalent information.


\begin{figure}
        \centering
        \includegraphics[width=\textwidth]{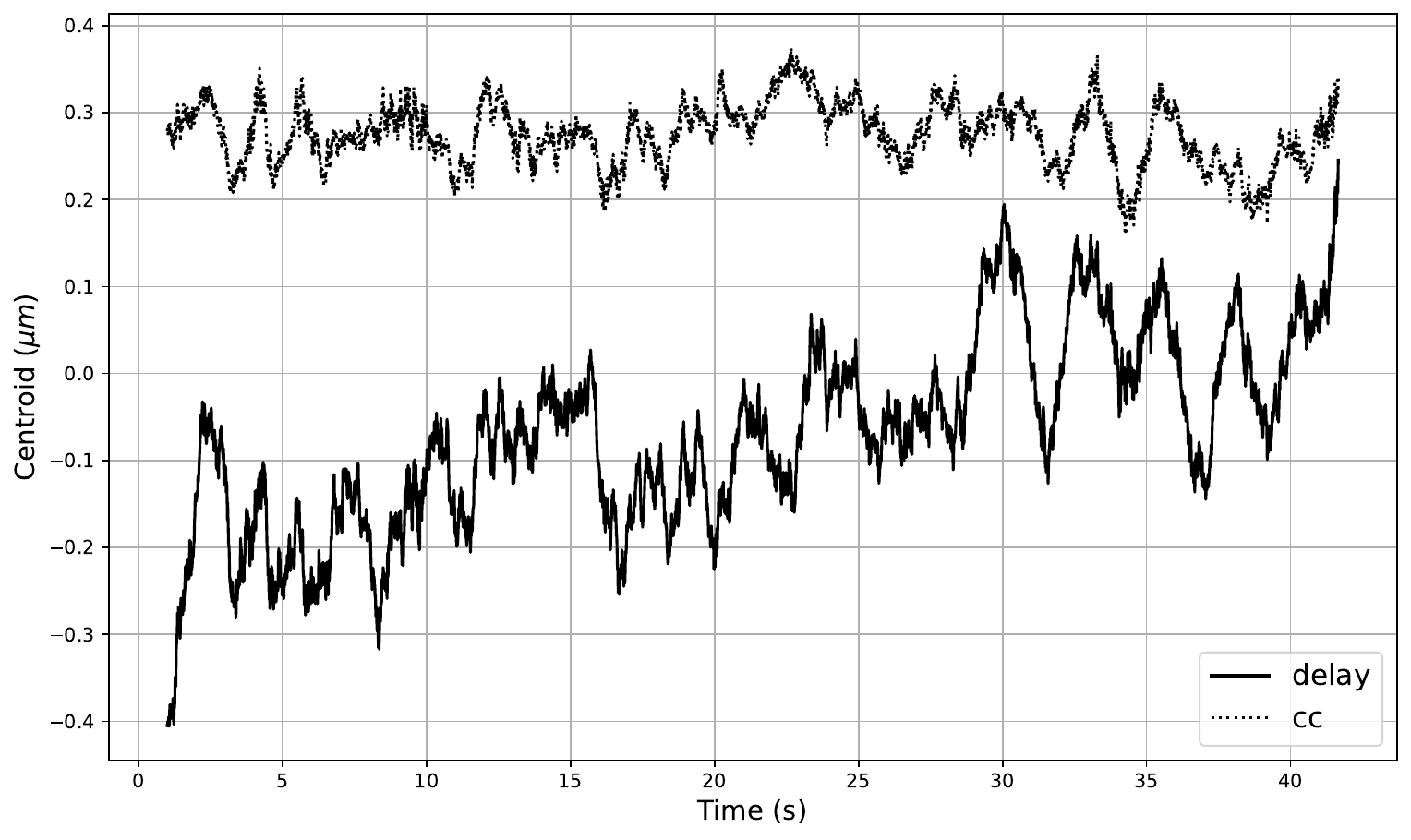}
        \caption{Visualization of the drift for the horizontal Beam Position Error in the Channel Cut (labeled as cc) and the Delay (labeled as delay) branches observed in an experiment using the HXRSND, without adjustment of any of the degrees of freedom. The rate of drift is observed to be $\approx$300~nm in 1 minute, and the standard deviation after subtracting the linear drift is 108~nm rms.}
        \label{fig:experimentaldriftrate}
\end{figure}

\subsection{Hyperparameter Selection}
For the Gaussian Process model used in this study, the inputs are the settings of the theta and chi knobs of the delay branch crystals, and the output is the predicted Beam Position Error. For the constrained TVBO experiment, an additional Gaussian Process model accepts inputs of the settings of the theta and chi knobs, and predicts if the Beam Intensity for this setting will adhere to the constraint. The parameters of the kernels for the Gaussian Process models are learned via training using gradient descent.
There are other parameters (termed hyperparameters) that are not learnt during model training, but have to be selected. These TVBO hyperparameters (specifically, the covariance kernel for the Gaussian Process model, the acquisition function for the Bayesian Optimization, the exploration parameter, the width of the sliding window, etc) were selected based on manual hyperparameter tuning. For the acquisition function, we tested different functions and selected the Upper Confidence Bound (UCB) acquisition function, with an exploration parameter value of $\beta=0.1$. After experiments, we selected a Matern kernel for the Gaussian Process model, with a smoothness parameter value of $\nu = \frac{5}{2}$. The width of the sliding window is an important parameter controlling the memory of the TVBO procedure. Selecting a window width that is significantly smaller than optimal leads to high variance and noise sensitivity in the optimization, as well as poor uncertainty estimates from the Gaussian Process model due to a dearth of samples to train on. Contrastingly, a width that is substantially larger than optimal leads to very slow adaption to temporal changes as older, less relevant data samples dilute the influence of recent and more relevant observations. Additionally, this adds to the computational budget as well, for the Gaussian Process model has to be trained at every iteration with all samples in the window. After experiments using the simulation, the width of the sliding window was selected as $w = 40$ samples. In this investigation, the focus is on linear drift observed in the HXRSND, however, for cases with non-linear drift a smaller sliding window width is preferred. The width of the sliding window is reflective of the persistence information content in past samples. In cases with linear drift, a larger sliding window enables better learning of the trend in the drift while minimizing the impact of noise in the samples on this estimation. In corresponding cases with non-linear drift, significantly older samples are less predictive of current behavior and can even bias the model incorrectly \cite{zhou2021no}, thus a smaller window width is preferred. The rate of drift was estimated from HXRSND operation data, as reported for one experiment in the dataset in Figure \ref{fig:experimentaldriftrate}, at $\approx$ 300~nm over $60$ samples (one minute). The aleatoric noise in the experimental data was calculated to be approximately 100~nm ($10\%$ of the stability goal) for the Beam Position Error measurements, and $1\%$ for the Beam Intensity measurements. These were incorporated in the data during the virtual experiments using the simulations.

\section{Drift Correction for HXRSND operation}
In this section, we outline virtual experiments utilizing simulations using TVBO for Drift Correction under different scenarios such as constant linear drift, drift with discontinuous jumps, and constrained TVBO for multi-objective control settings. The drift is introduced into the angular motion of the crystals, since the system performance relies most heavily upon the stability of these degrees of freedom.

\subsection{Continuous Linear Drift}

For this scenario, we assume the drift rate constant and equal in magnitude along all input features, along with a smaller stochastic component, $x_i (t) = x_i(0) + rt + \epsilon$. The direction of drift (positive or negative) is randomly assigned to each dimension. The goal of this study is to mimic small drifts due to thermal expansion of the system's constituent components on the minutes to hours timescale. The rate of drift (10~$\mu$m in 15 minutes) and the variance of the stochastic component (100~nm) are inferred from prior experimental data (see Figure~\ref{fig:experimentaldriftrate}). Our objective is to minimize the beam position error, and maintain this minimum over a given period of time as the system drifts.

\begin{figure}
\centering
\includegraphics[trim={10cm 4cm 8cm 3cm},clip, width=1.05\textwidth]{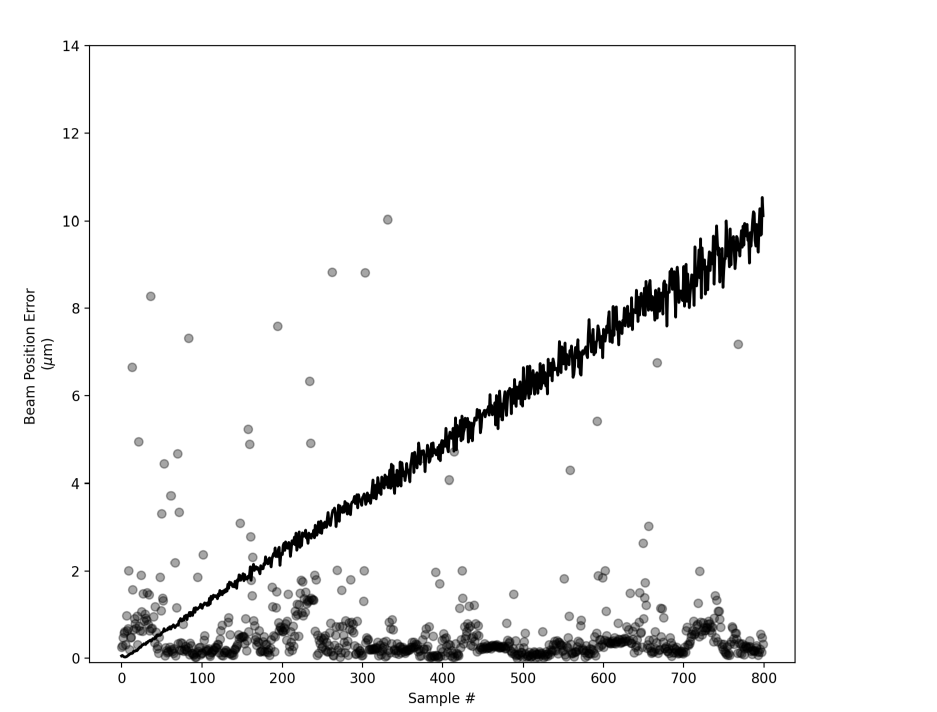}
\caption[]{Results using Time Varying Bayesian Optimization (TVBO) for Drift Correction with a constant, linear drift model, for the Beam Position Error in $\mu m$. The solid line reports the change in performance of the initial optimum setting due to drift. Each semi-transparent circle represents the Beam Position Error for a single sample generated using TVBO. The transparency in the circles is used to assist the reader visualize the location and density of the points even in areas of high overlap.
}
\label{fig:fig2}
\end{figure}

The TVBO results are outlined in Figure \ref{fig:fig2}. In our experiments, $40$ initial random samples were generated before commencing the TVBO. In Figure \ref{fig:fig2}, we show the results after the TVBO procedure is active and is controlling the setting to sample next. The samples generated by the TVBO procedure are reported using dark circles. The solid line reports the evolution of the Beam Position Error in the system if the initial optimum was retained. We observe that the TVBO procedure is able to maintain the Beam Position Error at low values. For the virtual experiment depicted in Figure \ref{fig:fig2}, the TVBO samples had a mean value of $ 0.72~\mu m$ for the Beam Position Error and a standard deviation of $1.82~\mu m$. This is further bolstered by statistics over $50$ runs of the experiment where over $80\%$ of the samples remain within the bound of $1~\mu m$ despite the drift.

\subsection{Discontinuous Drift}
In this scenario, we utilize a discontinuous rate of drift, simulating an experiment where drift correction is intermittent due to unavailability of parasitic measurements of the alignment status. During the initial experiment, no data is saved and the Drift Correction commences after the experiment. This leads to a jump in the drift, as is simulated in this experiment. Consequently, the drift can be mathematically expressed as 
\begin{equation*}
    x_i(t)=
    \begin{cases}
      x_i (t) = x_i(0), & \text{if}\ t \leq T \\
      x_i (t) = x_i(0) + rt + \epsilon, & \text{otherwise}.
    \end{cases}
\end{equation*}

Feedback loops for drift correction struggle with such sudden events for instance, a power supply interruption, that causes non-smooth drift. This affects the robustness of the drift correction.

\begin{figure}
\centering
\includegraphics[trim={10cm 4cm 8cm 3cm},clip, width=1\textwidth]{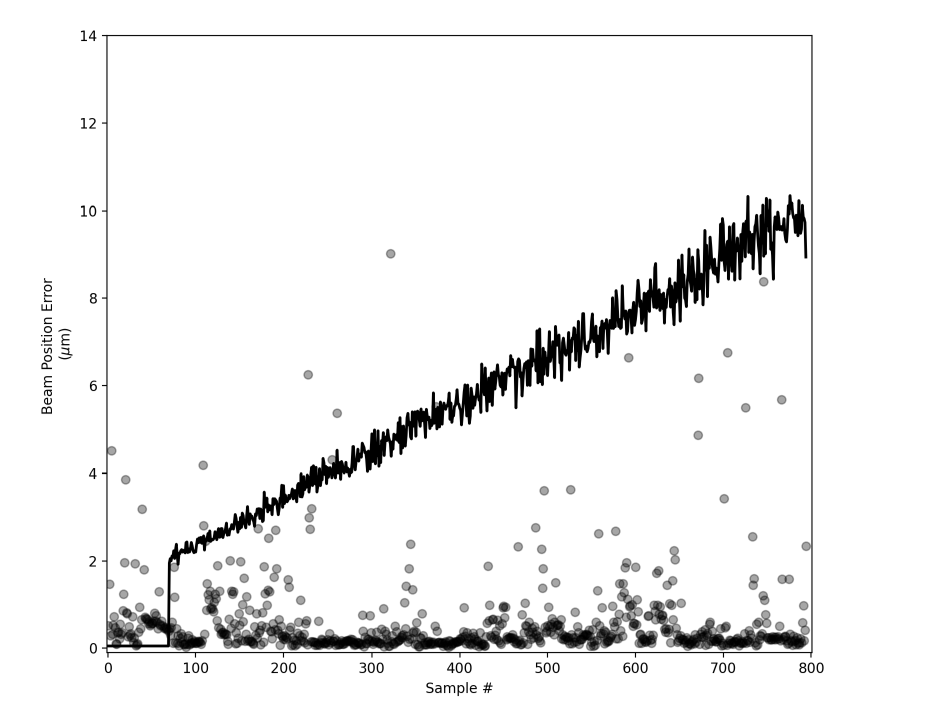}
\caption[]{Results using Time Varying Bayesian Optimization (TVBO) for Drift Correction with a Discontinuous, episodic drift model. The solid line reports the change in performance of the initial optimum setting due to drift. Each semi-transparent circle represents the Beam Position Error for a single sample generated using TVBO. 
}
\label{fig:fig3}
\end{figure}

The results for the experiment using TVBO are outlined in Figure \ref{fig:fig3}.  In the experiments, $70$ initial samples were generated before introducing the linear drift with a jump. The samples generated by TVBO are reported using dark circles. The solid line reports the evolution of the Beam Position Error in the system if the initial optimum was retained, where the discontinuity in the value of the Beam Position Error is reported. As is shown in Figure \ref{fig:fig2}, most of the samples generated by TVBO maintain a low value of the Beam Position Error. For the virtual experiment depicted in Figure \ref{fig:fig3}, the TVBO samples had a mean value of $ 0.88~\mu m$ for the Beam Position Error and a standard deviation of $2.04~\mu m$. Additionally, over $50$ runs of the virtual experiment, over $80\%$ of the samples remain within the bound of $1~\mu m$ despite the discontinuous drift.





\subsection{Drift Correction with Constraints}
In many real world scenarios, the process of optimization is not solely concerned with minimizing the objective function, but to also satisfy inequality constraints on additional outputs, thus defining feasible regions of the solution space. Constrained Bayesian Optimization extends standard Bayesian Optimization approaches to handle constraint functions, whose form may be unknown, by modeling both the objective and each constraint with separate Gaussian processes. The approach modifies standard acquisition functions to incorporate the probability of constraint satisfaction, using multiplicative terms such as $P(feasible|x)$, enabling the algorithm to balance exploration of feasible regions with exploitation for objective improvement. In the process of Drift Correction of the HXRSND, while minimizing the Beam Position Error, we are obligated to maintain the system throughput (i.e. Beam Intensity) at a high value. To this end, we carry out Constrained Time Varying Bayesian Optimization experiments, where we minimize the Beam Position Error as the objective, while treating the Beam Intensity value as an inequality constraint, where we maintain the Beam Intensity at $95\%$ of its initial set value. The set up for these experiments is identical to the Linear Drift case, but with the addition of the constraint upon the Beam Intensity.

\begin{sidewaysfigure}
\centering
\includegraphics[width=\textwidth]{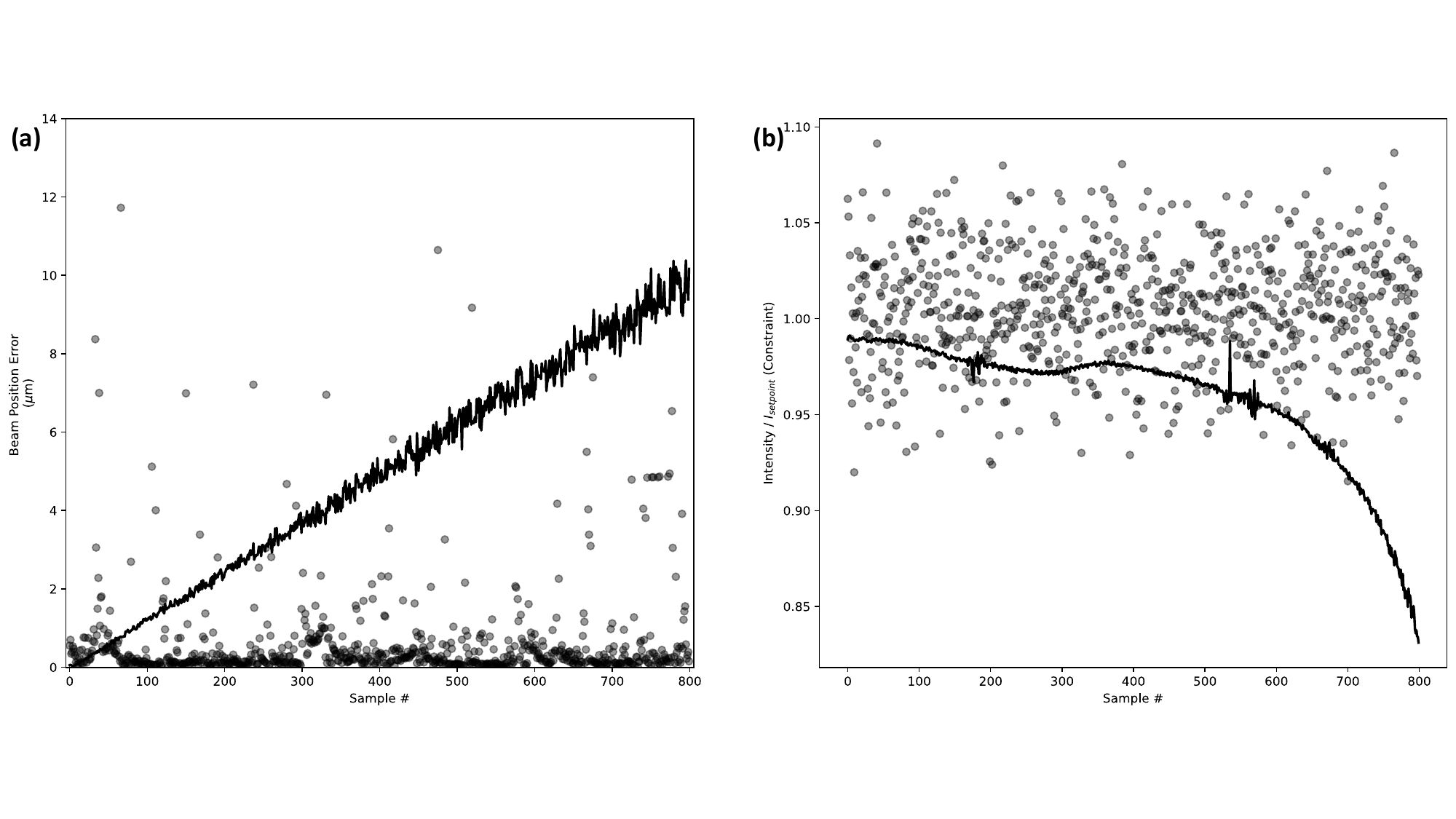}
\caption[]{Results using Constrained Time Varying Bayesian Optimization (TVBO) for Drift Correction. (a) The solid line reports the change in the Beam Position Error of the initial optimum setting due to drift. Each semi-transparent circle represents the Beam Position Error for a single sample generated using TVBO. (b) The solid line reports the change in the Beam Intensity of the initial optimum setting due to linear drift. Each semi-transparent circle represents the value of the constrained intensity generated by each sample generated using TVBO. $I_{setpoint}$ refers to the initial optimum value of the intensity. Thus, the intensity constraint is normalized by the initial optimum value of the intensity. For this virtual experiment using simulations, it was set to $\frac{I}{I_{setpoint}} > 0.95$.
}
\label{fig:fig4}
\end{sidewaysfigure}

The results for the experiment are outlined in Figure \ref{fig:fig4}. In Figure \ref{fig:fig4} (a), the Beam Position Error values using the TVBO procedure are contrasted against the change in the performance of the initial optimum setting. For the virtual experiment depicted in this figure, the TVBO samples had a mean value of $ 0.97~\mu m$ for the Beam Position Error and a standard deviation of $2.39~\mu m$. In Figure \ref{fig:fig4} (b), the Beam Intensity (used as the constraint) is visualized and contrasted against the the change in the performance of the initial optimum setting due to drift. The contours of the Beam Intensity is characteristic of crystal optics, characterized by a sharp rise followed by a gently sloping top leading to a sharp drop in value. This is observed in the change in the Beam Intensity of the initial setting due to drift. It is observed that almost all the samples generated by the TVBO procedure are able to maintain the value of the Beam Intensity above the specified constraint threshold. 





\section{Conclusions}
The utilization of the increasing brightness of modern light sources opens the opportunity for unprecedented insights into nanoscale and picosecond phenomena through increasingly sophisticated and complex experiments, but for the success of these experiments relies on beam stability with a tight tolerance over extended time periods. The primary hurdle to this end is temporal drift, from myriad and often unknown sources, in the system, that obligates frequent retuning after brief durations, resulting in reduced scientific throughput. In this study, we apply Time Varying Bayesian Optimization (TVBO) for temporal drift correction. This technique is applied to a complex optical Split And Delay system with a high dimensional input space, under different drift models representing real-life scenarios. It is exhibited that TVBO is able to account for and correct temporal drift, resulting in a stable beam. With the advent of stationary Bayesian Optimization for beam alignment from a cold start, TVBO may represent a convenient and proficient approach for drift correction. Using TVBO for drift correction can be adapted to additional X-ray beam conditioning and guidance systems, such as multi-crystal monochromators and grazing-incidence mirrors, to maintain sub-micron and nanoradian beam stability over the duration of experiments spanning several hours.

Considering the broader impact of this investigation, transitioning from human-in-the-loop drift correction to Time Varying Bayesian Optimization represents a paradigm shift from reactive stabilization to proactive, continuous performance optimization. Current drift correction approaches focus on Beam Position Error only, whereas TVBO continuously seeks an optimum in the presence of drift. In this regard, the beamline would not just be stable, but would attempt to operate at its peak achievable performance at all times, even as thermal, mechanical, and electronic conditions keep changing. Furthermore, TVBO can be used in multi-objective settings using different objective functions and constraints, which is not as simple while using feedback loops. These changes represent a step towards self-driving accelerator facilities, where control transitions from a system of manually-tuned, independent feedback loops to an automated agent that can re-optimize over the complex, high-dimensional and time-varying input space.

\begin{funding}
Use of the Linac Coherent Light Source (LCLS), SLAC National Accelerator Laboratory, is supported by the U.S. Department of Energy, Office of Science, Office of Basic Energy Sciences under Contract No. DE-AC02-76SF00515. This work was performed and partially supported by the US Department of Energy (DOE), Office of Science, Office of Basic Energy Sciences Data, Artificial Intelligence and Machine Learning at the DOE Scientific User Facilities program under the MLExchange Project (award No. 107514).
\end{funding}

\ConflictsOfInterest{The authors declare no conflicts of interest.
}

\DataAvailability{Data underlying the results presented in this paper are not publicly available at this time but may be obtained from the authors upon reasonable request.}

\bibliography{iucr} 

\end{document}